\documentclass[a4paper,11pt]{article}
\pdfoutput=1 

\usepackage{jinstpub} 

\title{First measurement of the surface tension of a liquid scintillator based on Linear Alkylbenzene (HYBLENE 113) }

\author[c]{J. Alt,}
\author[d]{D. Arutinov,}
\author[e]{O. Bezshyyko,}
\author[a]{T. Bretz,}
\author[b]{A. Brignoli,}
\author[b]{A. Conaboy,}
\author[f]{P. Deucher,}
\author[g] {F. De Paola,}
\author[g] {G. del Giudice,}
\author[g] {C. di Cristo,}
\author[g] {O. Fecarotta,}
\author[h] {A. Fiorillo,}
\author[c]{H. Fischer,}
\author[d]{H. Gl\"uckler,}
\author[d]{C. Grewing,}
\author[f]{A. Hollnagel,}
\author[b]{H. Lacker,}
\author[h] {A. Miano,}
\author[a,d]{G. Natour,}
\author[e]{V. Orlov,}
\author[h] {A. Prota,}
\author[a]{F. Rehbein,}
\author[b,1]{A. Reghunath,\note{Corresponding author.}}
\author[h] {A. Salzano,}
\author[d]{M. Schaaf,}
\author[b]{C. Scharf,}
\author[b]{J. Schmidt,}
\author[c]{M. Schumann,}
\author[b]{A. Vagts,}
\author[d]{S. van Waasen,}
\author[f]{M. Wurm}

\affiliation[a]{Physics Institute 3A, RWTH Aachen University, Sommerfeldstra\ss e 16, 52074 Aachen, Germany}
\affiliation[b]{Institut f\"ur Physik, Humboldt-Universit\"at zu Berlin, 12489 Berlin, Germany}
\affiliation[c]{Physikalisches Institut, Albert-Ludwigs-Universit\"at Freiburg, 79104 Freiburg, Germany}
\affiliation[d]{Forschungszentrum J\"ulich GmbH, Central Institute of Engineering, Electronics and Analytics, 52425 J\"ulich, Germany}
\affiliation[e]{Taras Shevchenko National University of Kyiv, Kyiv, Ukraine}
\affiliation[f]{Institut f\"ur Physik and PRISMA Cluster of Excellence, Johannes Gutenberg Universit\"at Mainz, Mainz, Germany}

\affiliation[g]{Department of Civil, Building and Environmental Engineering, Universit{\`a} di Napoli \lq \lq Federico II \rq \rq, Napoli, Italy}
\affiliation[h]{Department of Structures for Engineering and Architecture, Universit{\`a} di Napoli \lq \lq Federico II \rq \rq, Napoli, Italy}

\emailAdd{anupama.reghunath@physik.hu-berlin.de}

\abstract{We measured the surface tension of linear alkylbenzene (LAB) HYBLENE 113 mixed with Diphenyloxazole (PPO) as well as of pure LAB HYBLENE 113 as part of material studies for the liquid-scintillator based surround background tagger (SBT) in the proposed SHiP experiment. The measurement was performed using the iron wire method and the surface tension for linear alkyl benzene HYBLENE 113 plus PPO was found to be $(30.0\pm0.6)$ mN/m $22.0\pm 0.5$ \textdegree C and for pure HYBLENE 113, $(29.2\pm 0.6)$ mN/m at $21.0\pm 0.5$ \textdegree C.}

\collaboration[c]{(SHiP SBT collaboration)}

\begin{document}
\maketitle
\flushbottom
\section{Introduction}
\label{sec:intro}

In the proposed SHiP experiment \cite{SHiP:2015vad,SHIP:2021tpn} at the CERN SPS, the vacuum decay vessel has to be surrounded by a detector to detect muons entering the decay vessel from the beam-dump or from inelastic scattering events induced by neutrinos or muons in the decay vessel walls or its vicinity. The baseline detector concept for this surround background tagger (SBT) is based on cells filled with liquid scintillator, hermetically enclosing the lateral sides of the decay vessel which are readout by wavelength shifting optical modules (WOMs) \cite{Bastian-Querner:2021uqv}  equipped with silicon photo multipliers (SiPMs). The selected
liquid scintillator is linear alkyl benzene (LAB) together with 2.0\,g/l diphenyloxazole (PPO) as the fluorescent \cite{Ehlert_2019}. Linear alkyl benzene has been found to be a good candidate as solvent for liquid scintillation systems for other liquid scintillator experiments such as JUNO and SNO+ \cite{JUNO:2015sjr,SNO:2015wyx}.

A non-trivial task is to fill and empty the SBT, which consists of a large number of interconnected cells (O(2000)), with cell sizes of 80\,cm in the longitudinal direction, and between $\sim$80\,cm and $\sim$150\,cm in the azimuthal direction, depending on the location along the vacuum vessel. 
The behaviour of the liquid during filling and emptying depends on its surface tension, which is also an important input parameter to simulations of the filling/emptying process prior to vacuum vessel/SBT construction.
In particular, the surface tension influences the dimension and the shape of the gas bubbles dispersed into the fluid \cite{Han2010,Park2017}, with a subsequent influence on both the buoyancy and the drag acting on them. Thus, to correctly design the filling system of the SBT with the liquid scintillator, a precise assessment of the surface tension is required because of its influence on the velocity of the gas bubble into the fluid and their tendency to be pushed towards the gas-liquid interface for their expulsion. As LAB is widely used in liquid-scintillator detectors, the results of these measurements will be also beneficial in optimising the liquid handling systems of other experimental facilities.


Surface tension of a liquid is the result of inward and tangentially directed cohesive forces dominating over adhesive forces.
It results in a film-like behaviour in the liquid-air interfaces and is key to understand the wetting property of a liquid. In this study, we measure the surface tension of pure LAB HYBLENE 113 and of LAB HYBLENE 113\footnote{Hyblene 113 used in this study is from the manufacturer SASOL Limited.} mixed with PPO, which is currently in use as a liquid scintillator detector for building  SBT detector prototypes.
\section{Experimental Setup and Results}

\begin{figure}[]  \centering                       \includegraphics[width=0.4\textwidth]{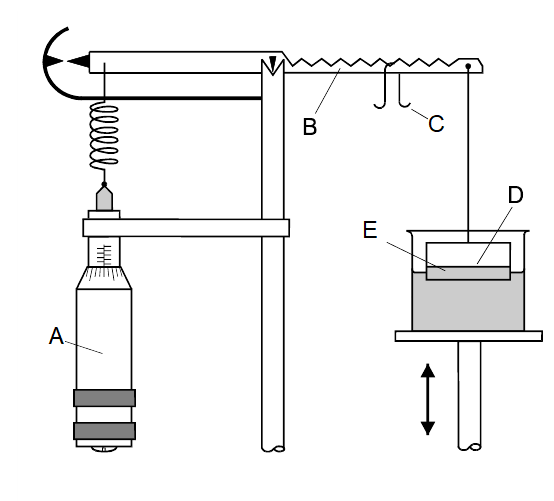}
    \caption{Experimental Setup for the Iron Wire method. In the figure, A, B and C denotes the measurement screw, balance beam and the calibration weight of the mechanical force compensation device. D corresponds to the iron wire, and E the lamella formed between the liquid and the iron wire. Image taken and modified with permission from  \cite{grundpraktikum}.
    } 
    \label{fig:exp_setup}
\end{figure}

The measurement of the surface tension is carried out using the wire frame method \cite{Gerthsen}. A thin wire of known dimensions is suspended on a mechanical force compensation device (see fig.\ref{fig:exp_setup}) consisting of a spring attached to a measurement screw. The wire is immersed in the liquid and is pulled upwards resulting in the formation of a lamella between the wire and the liquid surface. This thin layer of liquid pulls the wire downwards due to the surface tension, which can be measured by balancing the balance beam(B) with the measurement screw(A) on the spring.

The setup is initially calibrated using a known weight to calculate the spring constant. Once the calibration of the spring is completed, the position of equilibrium of the balance is read from the measurement screw. The wire is then lowered in the liquid until fully immersed. The beaker is slowly pulled downwards while simultaneously adjusting the measurement screw such that the balance is always maintained. At the point where the lamella breaks, the reading in the measurement screw is taken. The length difference in measured distances between the equilibrium point and the position at which the lamella breaks (denoted as $\mathit \Delta a$) is used for the calculation of the surface tension.

The surface tension is calculated as,
\begin{equation}
    \sigma=\frac{F_t}{2\cdot l}=\frac{1}{2\cdot l}\cdot \frac{\Delta a}{p}\cdot m\cdot g
\end{equation}
where $F_t$ is the force and $l$ is the length of the wire. $\Delta a$ is the difference in the measurements, $m$ is the mass of the calibration weight(C), $g$ is the gravitational acceleration, and $p$ is the spring constant. The factor 2 arises due to the formation of two films on either side of the wire. The length of the wire used for the experiment is $(50.00 \pm0.05)$ mm.

The corresponding uncertainty is

\begin{eqnarray}
  \Delta \sigma(x_i)&=& \left [
                \sum_{i} \left(\frac{\partial  \sigma}{\partial x_i}\cdot \Delta x_i \right)^2
                    \right ]^{\frac{1}{2}}
\end{eqnarray}

where $x_i\in \{l,\Delta a,p,m\}$

The error is dominated by the uncertainty arising from the measurement of $\mathit \Delta a$, and the combined contributions from other terms(eg. spring constant) accounts only for 10$\%$ of the total error. The main source of the error in $\mathit\Delta a$ comes from limitations in fine adjustment of the measurement screw during readings.


\begin{table}[htbp]
        \caption{Surface tension value of HYBLENE 113+PPO, HYBLENE 113, Denatured ethanol and 2-propanol taken using the iron wire method.  }
        \centering
        \begin{tabular}{ll|cc|ccc}
        \\
        \hline
        \hline
        \multicolumn{1}{c}{ } &&
        \multicolumn{1}{c}{Liquid Temperature (\textdegree C)} && 
        \multicolumn{3}{c}{Surface Tension Values(mN/m)} \\
        \hline
        && &&Results of the study && Literature\\
        \hline

        HYBLENE 113+PPO   &&22($\pm0.5$)&& 30.0($\pm0.6$) && \\
        HYBLENE 113 && 21($\pm0.5$)&&29.2($\pm0.6$) &&\\
        Denatured Ethanol   && 20($\pm0.5$)&& 22.0($\pm0.6$)&& 22.36$\pm0.01$*\\
        2-Propanol     &&20($\pm0.5$)&& 21.8($\pm0.5$) && 21.74**\;\\
    
        \hline 
        \multicolumn{7}{l}{\small 
        *\cite{GONCALVES20101039}}\\
        \multicolumn{1}{l}{\small 
        **\cite{Vazquez1995}
        $\pm 0.4$  \% error at 20\textdegree C.}\\
        \hline 
        \hline
  
        \end{tabular}
        \label{tab:par}
    \end{table}

The measured value for the surface tension for the LAB: HYBLENE 113+PPO is $(30.0\pm 0.6)$ mN/m at $22.0\pm 0.5$ \textdegree C and for HYBLENE 113 alone is $(29.2\pm 0.6)$ mN/m at $21.0\pm 0.5$ \textdegree C. The two values
agree within uncertainties, which is not unexpected given the small concentration of PPO. Hence, it is expected
that the surface tension of LAB-based liquid scintillators mixed with other fluors than PPO has values very 
close to the ones reported here.
To verify our experimental setup, surface tension of denatured ethanol and 2-propanol were measured and were found to be in good agreement with the literature values (see Table \ref{tab:par}).

\acknowledgments This work is funded by the Deutsche Forschungsgemeinschaft (Germany).

\end{document}